# Attosecond screening dynamics mediated by electron-localization

M. Volkov,[1*] S. A. Sato,[2] F. Schlaepfer,[1] L. Kasmi,[1] N. Hartmann,[1] M. Lucchini,[1†] L. Gallmann,[1] A. Rubio,[2,3] U. Keller[1]

[1]Department of Physics, ETH Zurich, 8093 Zurich, Switzerland.

[2]Max Planck Institute for the Structure and Dynamics of Matter and Center for Free-Electron Laser Science, Luruper Chaussee 149, 22761 Hamburg, Germany

[3]Center for Computational Quantum Physics (CCQ), The Flatiron Institute, 162 Fifth avenue, New York NY 10010, USA

*Correspondence to: volkovm@phys.ethz.ch

†Current address: Department of Physics, Politecnico di Milano, 20133 Milano, Italy

**Transition metals with their densely confined and strongly coupled valence electrons are key constituents of many materials with unconventional properties[1], such as high-Tc superconductors, Mott insulators and transition-metal dichalcogenides[2]. Strong electron interaction offers a fast and efficient lever to manipulate their properties with light, creating promising potential for next-generation electronics[3–6]. However, the underlying dynamics is a fast and intricate interplay of polarization and screening effects, which is poorly understood. It is hidden below the femtosecond timescale of electronic thermalization, which follows the light-induced excitation[7]. Here, we investigate the many-body electron dynamics in transition metals before thermalization sets in. We combine the sensitivity of intra-shell transitions to screening effects[8] with attosecond time resolution to uncover the interplay of photo-absorption and screening. First-principles time-dependent calculations allow us to assign our experimental observations to ultrafast electronic localization on d-orbitals. The latter modifies the whole electronic structure as well as the collective dynamic response of the system on a timescale much faster than the light-field cycle. Our results demonstrate a possibility for steering the electronic properties of solids prior to electron thermalization, suggesting that the ultimate speed of electronic phase transitions is limited only by the duration of the controlling laser pulse. Furthermore, external control of the local electronic density serves as a fine tool for testing state-of-the art models of electron-electron interactions. We anticipate our study to facilitate further investigations of electronic phase transitions, laser-metal interactions and photo-absorption in correlated electron systems on its natural timescale.**

A characteristic thermalization time of laser-excited hot electrons in solids is in the femtosecond regime and becomes faster with stronger electron interactions[7]. Therefore, attosecond time resolution is required to resolve coupled electron dynamics during the laser-matter interaction before electron thermalization has occurred. Attosecond transient absorption spectroscopy revealed electric-field-guided electron dynamics in simple dielectrics and semiconductors[9–14]. However, transient absorption studies of localized and strongly interacting electrons, such as on d- and f- orbitals of transition metals, have been limited to the few-femtosecond regime[15]. Transition metal elements are the key constituents of many materials exhibiting remarkable properties, such as Mott insulators, high-Tc superconductors and transition metal dichalcogenides[2]. Understanding the coupled-electron dynamics in these systems is central for their applications in optoelectronics, energy-efficient electronics, magnetic-memory devices, spintronics and new solar cells[16]. Transition metal elements were studied with attosecond photoemission spectroscopy[17–19]. However, state-of-the-art theoretical treatment[20] of photoemission from metals on the attosecond timescale neglects any electronic perturbation induced by the optical field beyond the skin-effect. This approximation is based on efficient screening of the optical field by surface electrons[21] and space-charge limitation of the maximum optical field intensity[22]. In contrast, femto- and picosecond core-level photo-absorption of metals systematically shows signatures of excitation-induced electronic structure modification at high intensities[23–26].

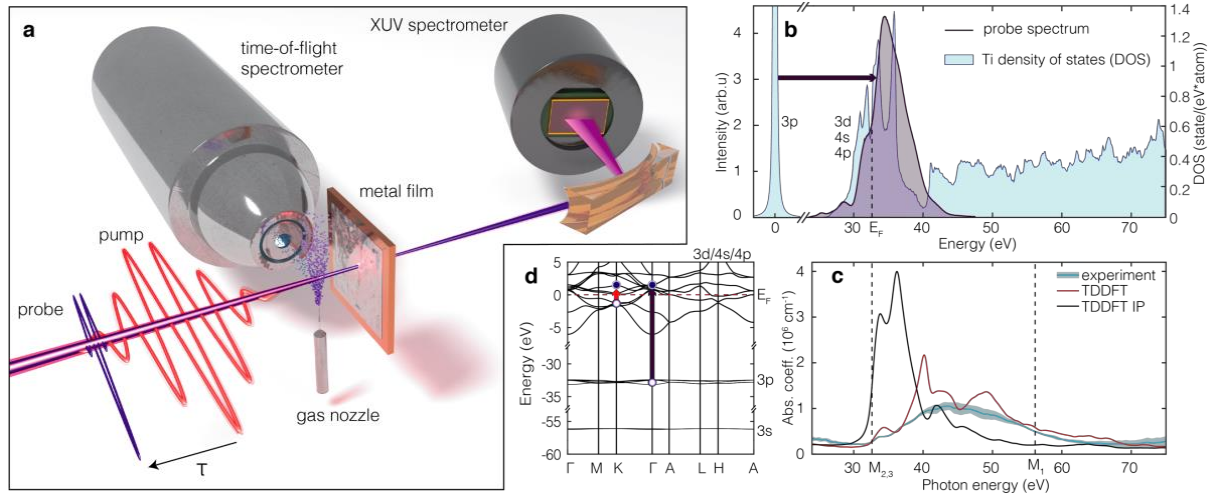

**Fig. 1 | Attosecond transient absorption spectroscopy. a**, Schematic of experimental apparatus for acquisition of photon and photoelectron spectra to measure both the incoming transient near-infrared (NIR) pump laser field and the spectrally resolved probe absorption in the extreme-ultraviolet (XUV). Photoionized electrons from the Ne gas nozzle are accelerated by the optical field and collected with the time-of-flight spectrometer. Transmitted probe radiation is recorded with an XUV spectrometer. **b**, Calculated electronic density of states of Ti (blue area). Violet area shows the probe spectrum with respect to 3p level. The vertical dashed line indicates the Fermi energy position. **c**, Experimental absorption spectrum of titanium (blue line, shaded area shows measurement uncertainty, see SI) compared to TDDFT calculations. The red line shows the absorption coefficient from a full simulation, whereas the black line shows the result of the independent-particle model. **d**, Calculated band structure of Ti. Red and violet arrows indicate the pump and probe transitions, respectively.

To bridge this gap of intensities and timescales, we used attosecond transient absorption spectroscopy to study collective electron dynamics in the transition metals Ti and Zr. In the following we discuss the phenomena observed in Ti, while physically analogous behavior of Zr is presented in the supplementary information (SI). A comprehensive description of our experimental apparatus is given in[12,27]. In the pump-probe experiments (Fig. 1a), few-cycle near-infrared (NIR) laser pulses excite electrons in a 50-nm-thick, free-standing Ti film. The induced change in the sample transmission around the $M_{2,3}$ absorption edge is probed by single attosecond pulses (SAP) in the extreme-ultraviolet range (XUV, ~35 eV, Fig. 1b) with a pulse duration of 260 as. The attosecond pulses are produced via high-harmonic generation in Kr. In addition, a Ne gas jet is placed in front of the metal film sample. Photoelectrons from the jet are collected with a time-of-flight spectrometer in a streaking experiment, providing precise information on the electric field of the NIR pump pulse.

Figure 1c shows the absorption spectrum of Ti, obtained with attosecond pulse trains covering the photon energy range from 25 to 75 eV (SI). We compare it with *ab-initio* calculations based on the time-dependent density functional theory (TDDFT) in the local density approximation (LDA), for a hexagonal close-packed (hcp) Ti cell, see SI for the details. Note that this theoretical framework has been shown to accurately describe electronic screening effects[28] as well as attosecond electron dynamics in solids[10,12,29]. The full calculation (shown in red in Fig. 1c) closely reproduces the experimental spectrum, whereas the independent particle approximation (Fig. 1c, black), which ignores electron-electron interactions (see SI for definition), fails to describe it. The underlying many-body process is known as the local field effect[30], ascribed to the additional potential induced locally by charge displacement. The Ti $M_{2,3}$ absorption edge, located at 32.6 eV[31], is dominated by the intra-shell 3p-3d electronic transition (Fig. 1d), featuring a large overlap of initial- and final-state wavefunctions[8]. The $M_1$ absorption edge at 56.1 eV caused by 3s-to-4p transitions is much less prominent due to the delocalized character of 4p final states. Strong interaction of electrons, participating in the transition and screening processes leads to a collective excitation, which manifests as a giant resonance[32] above the $M_{2,3}$ absorption edge. Therefore, our probing mechanism is sensitive to the dynamics of screening and collective electron motion[8,33] via many-body effects. In contrast, a conventional probing mechanism is usually based on the independent-particle picture and only senses the final-state occupation. We note that the experimental data in Fig. 1c exhibits no prominent peaks in the region of 32 to 52 eV, in contrast to the TDDFT calculations (red curve in Fig. 1c). We suggest that the discrepancy originates from the smoothening role of Auger relaxation, which is not included in the calculation.

We record the transmitted probe spectra with pump on and off in a fast sequence, using a mechanical shutter, and calculate the induced optical density (ΔOD, eq. S1.1, SI). Figure 2a shows the pump-induced change of absorption as a function of pump-probe delay in a 20-fs window with 250-as delay steps. The pump peak intensity is $I_{pump} = 7.5 \pm 0.7 \times 10^{11}$ W/cm$^2$ and the center photon energy is 1.55 eV. In the presented data, we have subtracted the delay-independent thermal background (see SI).

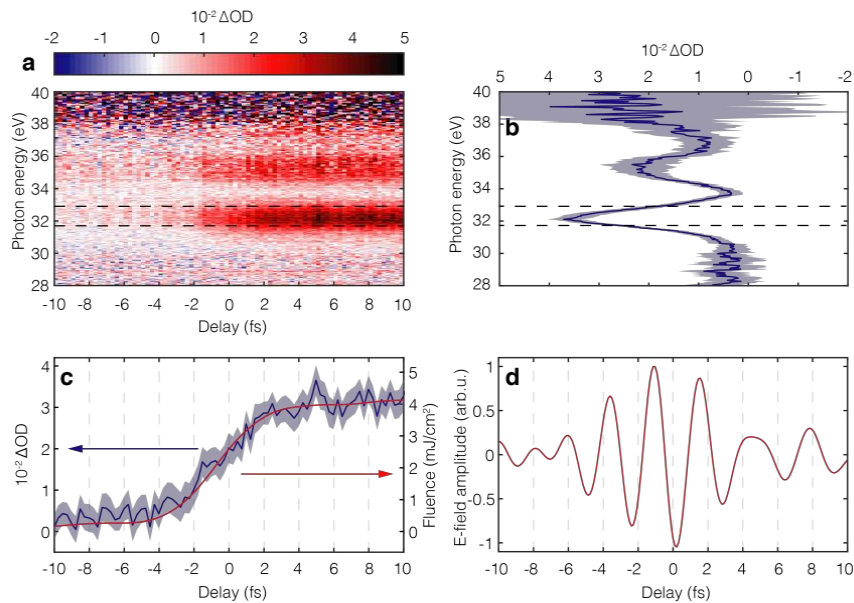

**Fig. 2 | Experimental pump-probe results. a**, Transient absorption spectrogram showing pump-induced optical density change ΔOD, probed with a single attosecond pulse (SAP). Positive delays stand for the XUV probe arriving after the NIR pump pulse. **b**, Profile of the induced optical density at positive delays (0 to 10 fs). **c**, Delay-dependent signal extracted from the data in (**a**) in the photon energy band of 31.7 to 32.9 eV (dashed horizontal lines in panels **a**, **b**) with incident pump fluence reconstructed from the streaking experiment (red curve). **d**, Electric field of the NIR pump pulse retrieved from the streaking experiment and associated to the fluence in panel (**c**).

The pump-probe spectrogram (Fig. 2a) reveals an optical density profile that rapidly increases in amplitude and exhibits several peaks along the energy axis (Fig. 2b). Previous studies of attosecond electron dynamics in semiconductors and dielectrics revealed field-driven, oscillatory dynamics[9–14]. In contrast, in the metallic system of titanium, we observe an instantaneous, linear response. Figure 2c shows that the optical density transient follows the laser fluence extracted from the streaking experiment (Fig. 2d). By analyzing several independent experimental data sets with a linear convolution model, we estimate an upper limit of the microscopic response time to the pump fluence as 1.1 fs with the experimental accuracy of 290 as (see SI). Although the time-evolution of the signal is guided by pump absorption, the spectrogram does not display any induced transparency features (Fig. 2a, blue false color), indicating that the expected state-filling effect is completely cancelled by a many-body phenomenon.

To provide a more detailed understanding of the underlying electronic dynamical rearrangement and its effect on Ti absorption, we simulate the pump-probe experiments with *ab-initio* TDDFT calculations[12,14,34]. Figure 3a (red line) displays the absorption change of hcp Ti, induced by a 10-fs NIR pump pulse with center photon energy of 1.55 eV and incident intensity of $I_{pump} = 1\times10^{12}$ W/cm$^2$. Our simulation successfully reproduces the main absorption feature around 32 eV in highly non-equilibrated state of Ti. We compare these results with electron-thermalized state of Ti and find that a particular excited electron distribution has little influence on the induced spectral feature (Fig. 3a, 3c). Accordingly, experiments show that XUV absorption at the end of the pump pulse and 12 fs later differs only slightly, hence thermalization has little influence on our observables (Fig. 3b). The independent-particle (IP) model (Fig. 3a, black), which ignores electron-electron interactions, merely reflects the electron-hole population (Fig. 3c) and exhibits transparency above the $M_{2,3}$ edge. In contrast, induced transparency is suppressed in the full model, in agreement with the experiment.

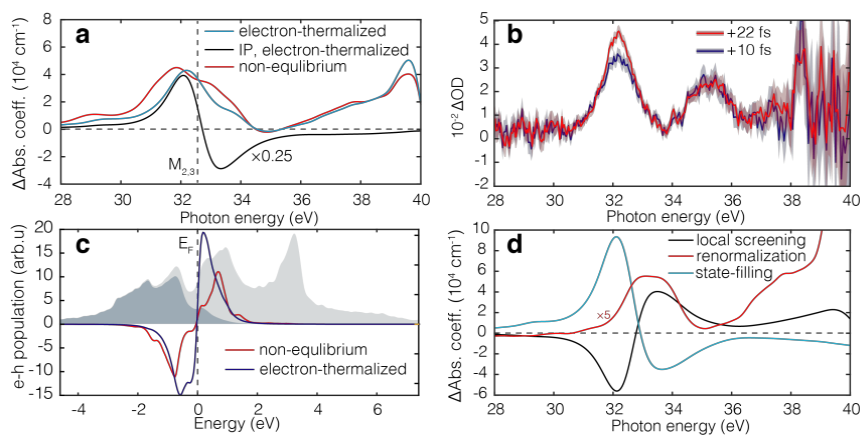

**Fig. 3 | *Ab-initio* calculation results. a**, Calculated change of absorption coefficient in non-equilibrium, immediately after excitation with incident pump intensity of $1\times10^{12}$ W/cm$^2$ (red line). Thermo-induced absorption change with electron-temperature of 0.315 eV from a full model (blue line) and independent-particle (IP) approximation (black line, scaled down by a factor of 0.25). **b**, Experimental optical density profiles 10 fs and 22 fs after the pump-probe overlap (blue and red lines). **c**, Non-equilibrium electron-hole population after the NIR pump excitation (red line), and corresponding thermalized distribution (blue line). Light-blue and dim-blue areas show the electronic density-of-states and electron population in the electron-thermalized phase, respectively. **d**, Decomposition of the induced absorption coefficient for a thermalized system into three physical components: state filling (blue line), electronic structure renormalization (red line, scaled up by a factor of 5) and local-screening modification (black line).

To gain further insight, we separate the TDDFT calculation into three physical factors: electronic structure renormalization, state-filling and local-screening modification (see SI). Figure 3d shows that the effect of state-filling (blue line) and the local-screening modification (black line) dominates the absorbance change. Importantly, the induced transparency above the $M_{2,3}$ edge due to the state-filling effect is primarily canceled by the local screening modification. As a result, the single positive peak around the absorption edge is formed, indicating a breakdown of the independent particle picture by the strong screening of localized *d*-orbitals. The renormalization effect is well-approximated by a red-shift of the absorption edge (see SI). This fact indicates enhancement of the static screening around the Ti atom due to increased electron-localization, in agreement with previous calculations of thermo-induced renormalization[35]. Finally, modified electron interaction also influences the core-hole lifetime, which is determined by super-Coster-Kronig Auger decay, involving two interacting 3d electrons. An increase of d-electron population thus shortens the semicore-hole lifetime. We find that additional empirical broadening of 0.15 eV qualitatively reproduces the second, weaker absorption maximum (35.5 eV, see Fig. 3b), though slightly shifted to higher energies (Fig. S10), thus capturing the essential features of experimental absorption spectrum.

Based on the agreement between theory and experiments we propose that the pump absorption results in an ultrafast modification of screening in the electronic system of Ti. In additional support of our hypothesis, we investigate in detail the underlying microscopic dynamics in real time and real space. Figure 4 shows the TDDFT simulation results of a pump field interacting with an hcp Ti cell. The laser electric field is parallel to the c-axis of the Ti crystal (Fig. 4a), and we visualize the electronic density transients in the c/a plane.

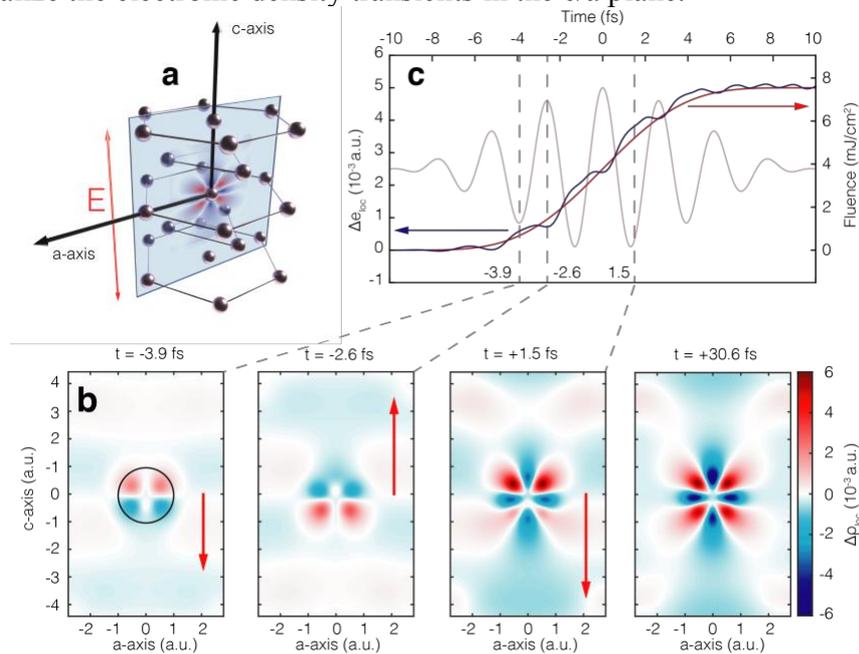

**Fig. 4 | Real-time and real-space theoretical investigation of the laser-driven microscopic electron dynamics in Ti. a**, Simulated laser field is parallel to the c-axis of the hexagonal-close-packed (hcp) Ti cell, blue plane shows the c/a-axis cut. **b**, Electron density dynamics in the c/a plane during the laser cycle, displayed in false color (red for increased electron density). **c**, Localized charge around Ti atom (blue line), closely following the laser fluence (red line). Localization region is marked with a black circle in panel (**b**).

Figure 4b shows the electron density difference induced by the NIR pump pulse. On top of the field-induced polarization dynamics, we observe a buildup of charge density, acquiring a distinct d-orbital shape by the end of the pulse. Therefore, the pump creates an inflow of electronic density towards the Ti atom and localizes it on the d-orbitals. To illustrate this, we integrate the localized electron density around a Ti atom and plot it as a function of time along with the pump fluence (Fig. 4c). As seen from the figure, electron localization closely follows the laser fluence in time, similar to the peak amplitude of the experimental transient absorption (Fig. 2c). This supports our hypothesis, that the transient absorption peak originates from ultrafast electron localization due to the pump pulse. The NIR pump modifies the atomic screening resulting in the absorption peak structure around the $M_{2,3}$ absorption edge, as discussed.

We note that the observed transient absorption features are robust against variation of experimental parameters, such as sample thickness, surface oxidation state and pump photon energy (see SI). Additionally, TDDFT calculations show no significant dependence on the crystal structure. We also observe identical induced spectral features at the $N_{2,3}$ absorption edge of Zirconium, suggesting a universal character of this phenomenon for transition metals (see SI). The possibility of charge localization in the metallic system of Ti is based on the relatively inefficient screening by 4s/4p electrons, since their wavefunctions weakly penetrate the 3d spatial region[36]. The discussed effect has essentially an atomic origin, suggesting that a photo-induced change of electron localization must also govern the first step of the interaction of light with transition metal compounds. Localization of valence electrons is expected to vary macroscopic properties of transition metals, such as conductivity, reflectivity, plasmonic behavior, magnetization and catalytic activity at surfaces. A particularly high electron density in transition metals, associated with fastest screening rates in the valence band, allows therefore to manipulate these macroscopic properties at unprecedented rates, in agreement with recent studies[37,38]. Our results unambiguously demonstrate photo-manipulation of screening in transition metals via ultrafast electron localization below the electron thermalization timescale.

**Supplementary Information** is available in the online version of the paper

**Acknowledgments:** We thank Prof. E. Krasovskii for valuable discussions. S.A.S. and A.R. thank M.J.T. Oliveira for helping with the generation of a transferable pseudopotential for Ti dealing with semicore electrons. This work was supported by the National Center of Competence in Research Molecular Ultrafast Science and Technology (NCCR MUST) funded by the Swiss National Science Foundation. We acknowledge financial support from the European Research Council (ERC-2015-AdG-694097), and the European Union's Horizon 2020 Research and Innovation program under Grant Agreements no. 676580 (NOMAD). S. A. S. acknowledges support by Alexander von Humboldt Foundation.

**Author Contributions:** M.V., S.A.S., L.G., A.R. and U.K. supervised the study. M.V., F.S., N.H, L.K. and M.L. conducted the experiments. M.V. and F.S. analysed the experimental data. S.A.S. and A.R. developed the theoretical modelling. All authors were involved in the interpretation and contributed to the final manuscript.

**Author Information:** Correspondence and requests for materials should be addressed to M.V. (volkovm@phys.ethz.ch) and U.K. (keller@phys.ethz.ch).

# Supplementary Information

M. Volkov,[1]* S. A. Sato,[2] F. Schlaepfer,[1] L. Kasmi,[1] N. Hartmann,[1] M. Lucchini,[1]†
L. Gallmann,[1] A. Rubio,[2,3] U. Keller[1]

[1]Department of Physics, ETH Zurich, 8093 Zurich, Switzerland.

[2]Max Planck Institute for the Structure and Dynamics of Matter and Center for Free-Electron Laser Science, Luruper Chaussee 149, 22761 Hamburg, Germany

[3]Center for Computational Quantum Physics (CCQ), The Flatiron Institute, 162 Fifth avenue, New York NY 10010, USA

*Correspondence to: volkovm@phys.ethz.ch

†Current address: Department of Physics, Politecnico di Milano, 20133 Milano, Italy

## S1. Supplementary methods

### Experimental apparatus and measurement procedure

The pump-probe experiments were carried out on the attosecond setup that is extensively described in refs.[12,27]. A commercial Ti:sapphire amplifier system followed by a double-filament compression setup delivers few-cycle (≈ 5 to 7 fs), CEP-stabilized ≈ 350 μJ pulses at a center wavelength of 780 nm and at 1 kHz repetition rate. A beam splitter separates 20% of the pulse energy and sends it to a piezo-actuated delay line, while the remaining 80% is used to generate extreme-ultraviolet (XUV) probe pulses. The generation is achieved by sending the beam through polarization-gating optics and focusing onto a 3-mm long cell filled with a noble gas (Ar, Kr or Xe). The polarization-gating[39] setup allows us to switch between the generation of single attosecond pulses (SAPs) and attosecond pulse trains (APTs). The XUV pulses are further separated from the HHG-driving near-infrared (NIR) radiation with a 100-nm thick Al foil and collinearly recombined with the delayed pump beam by a center-hole mirror. The delay between the two interferometer arms from the beam splitter up to the recombination mirror is actively stabilized by introducing a weak co-propagating continuous-wave (CW) 473–nm laser beam. A fast feedback loop acts on the piezo delay actuator, stabilizing the pump-probe delay to a desired value with 15 as rms uncertainty. A mechanical shutter in the pump arm of the interferometer is used for data acquisition with or without pump in a fast sequence. We define the pump-induced optical density as follows:

$$\Delta OD = ln \frac{I_{pr}^{no\ pump}}{I_{pr}^{pump}(\tau)} \tag{S1.1}$$

where $I_{pr}^{pump}$, $I_{pr}^{no\ pump}$ are the transmitted probe spectral intensities with and without pump radiation, $\tau$ is the pump-probe delay with positive values for the probe arriving after the pump.

### Sample characterization

Commercial polycrystalline metal foils from two different suppliers are mounted on copper TEM-grids with 9 windows. The unsupported window areas are 300 μm in diameter (Fig. S1A). The pump/probe spot sizes on the sample are ~ 50 μm. Bare metal foils of 50-nm thick Ti, 100-nm thick Ti and 100-nm thick Zr are provided by Lebow Co[40]. Passivated 48.6-nm thick Ti films (Luxel Co.[41]) were covered with 4.9 nm of amorphous carbon on both sides of the film without breaking the vacuum. We have carefully characterized the sample absorption spectra before and after exposure to pump pulses. The absorption spectrum of a bare 50-nm thick Ti foil is in excellent agreement with a free electron laser/synchrotron measurement of nominally identical samples from the same supplier[25]. The film thickness is determined from a physical measurement, but the foil wrinkles may add a geometrical uncertainty for the effective optical density. We have found that the best fit using the CXRO database values[42] is achieved for the film thickness 15% larger than nominal (55.9 nm for Ti-passivated and 57.5 for Ti-oxidized). We also observe a noticeable difference in the static absorption between passivated (Fig. S1B) and oxidized (Fig. S1C) samples. The thickness and composition of the Ti surface oxide layer formed under ambient conditions is taken from literature[43] and agrees well with the measured absorption of bare oxidized Ti foils. We note that the magnitude of the giant resonance of Ti at 45 eV significantly varies within the available literature [42,44,45].

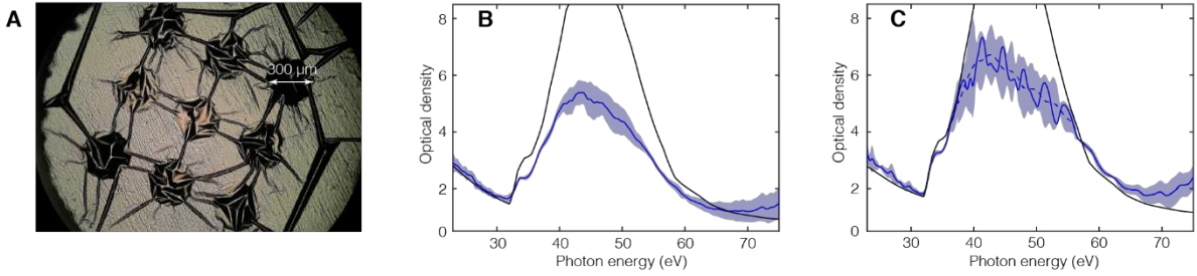

**Fig. S1.** **(A)** Microscope image of the Ti target **(B)** Blue line: experimentally determined optical density of carbon-passivated Ti foils. The blue-shaded area shows the standard deviation. Black line: optical density for 55.9-nm thick Ti and 2 4.9-nm thick layers of C according to the CXRO database[42]. **(C)** Same for bare oxidized Ti foils. Here, the black line shows the optical density of 57.5 nm of Ti with 2x(2 nm $Ti_2O_3$, 2 nm $TiO_2$, 2 nm TiO) for the oxide layer. The blue dashed line shows a smoothed average optical density in the highly-absorbing region to suppress artefacts due to the discrete nature of the APT spectrum.

In order to characterize the absorption of our samples in a broad energy region, we generated APTs in Xe, Kr and Ar. By changing the double-filament compression parameters we are able to tune the APT spectrum to close the gaps between harmonics (Fig S2).

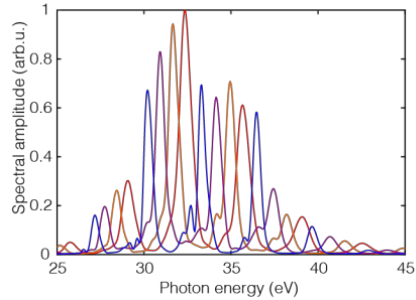

**Fig. S2**. Four examples of APT spectra, shifted by tuning the nonlinear compression of the HHG-driving pulse

In the static absorption measurement procedure, we acquire 10 pairs of signal (sample in) and reference (sample out) spectra in sequence by moving the motorized sample manipulator. Each signal/reference spectrum is an accumulation of ~$10^4$ XUV pulse trains, acquired rapidly in ~10 s. We then calculate the total optical density, $\langle \alpha \rangle$, as a weighted average over different probing spectra:

$$\langle \alpha \rangle = \frac{\sum \alpha_i w_i}{\frac{n-1}{n}\sum w_i}, \qquad w_i = \frac{1}{\sigma_{tot\, i}^2} \qquad (S1.2)$$

for each shape of the probe $i$ (Fig. S2) and corresponding absorption spectrum $\alpha_i$ the variance comprises of fast and slow fluctuations:

$$\sigma_{tot\, i}^2 = \langle \sigma_{fast\, j}^2 \rangle_i + \sigma_{slow\, i}^2. \qquad (S1.3)$$

The fast fluctuation term reflects the CEP and laser intensity fluctuation on a timescale of tens of milliseconds,

$$\sigma_{fast\,j}^2 = \left(\sigma_j^{I\,ref}/I_j^{ref}\right)^2 + \left(\sigma_j^{I\,sig}/I_j^{sig}\right)^2 \tag{S1.4}$$

where $j \in [1,10]$ is the index for each signal/reference pair, and $\sigma_j^I$ is the standard deviation over $10^4$ laser shots. The slow fluctuation term accounts for drifts between subsequent sample insertions (timescale of tens of seconds) and takes into account the position uncertainty of wrinkled metal foils:

$$\sigma_{slow\,i}^2 = \frac{\sum_j \left(\alpha_{j,i} - \langle\alpha\rangle_i\right)^2}{N-1}, \tag{S1.5}$$

where $\alpha_{j,i}$ is the optical density for each of the 10 signal/reference pairs, taken with *i-th* probe spectrum. The final uncertainty of the optical density is estimated as:

$$\sigma_\alpha^2 = \frac{\sum(\alpha_i - \langle\alpha\rangle)^2 w_i}{\frac{n-1}{n}\sum w_i}. \tag{S1.6}$$

**Thermal background and film thickness**

In the course of the pump-probe delay scan the samples are exposed to an average power of ~0.5 mW, resulting in elevated temperature. This effect manifests as a delay-independent background, which is separated from the transient signal at negative pump-probe delays. Its presence implies that the thermal background lifetime is longer than 1 ms, the inverse laser repetition rate. By reducing the repetition rate of our laser system from 1 kHz to 200 Hz (Fig. S4A) with a mechanical chopper, the amplitude of the background signal is reduced by half. We therefore estimate its lifetime to 5 ms. At the same time, reducing the repetition rate does not affect the transient part of the signal observed at positive delays (Fig. S4B).

We emphasize, that the thermo-induced spectral features are fully reversible, i.e. the sample completely recovers its static absorption after a few milliseconds. This condition holds within the intensity range of our pump-probe experiments. Only above the threshold intensity of $I_{pump}>10^{12}$ W/cm$^2$, close to the material breakdown, we detect a permanent decrease in the sample transmission after exposure to the pump radiation.

Further, we investigate the dependence of the thermal background on the sample thickness. As shown in Fig. S4C, the background amplitude is proportional to sample thickness, and therefore, originates from the bulk of the sample. The transient part, on the other hand, does not show any significant dependence on the thickness (Fig. S4D). The latter is expected, since the skin-depth of Ti is around 13 nm at 780 nm[46].

Although the ground-state of Ti at standard conditions has a hexagonal close-packed (hcp) structure, a number of studies showed the presence of a face-centered cubic (fcc) phase in thin films [47]. Moreover, an hcp-fcc transition may be caused by increased temperature[48]. We calculated the difference of the absorption coefficients for hcp and fcc Titanium (Fig. S4E). Although the change in fcc/hcp absorption resembles the thermal background profile, the total optical density change is one order of magnitude larger than the experimentally observed thermal background. We conclude, that if a hcp-to-fcc transition takes place, it only affects 10% of the material. Moreover, finite-temperature calculations show that both phases, fcc and hcp, have similar transient responses in the optical density (Fig. S4F). This supports the generality of our conclusions about the nature of the 32-eV peak being a renormalization of a collective excitation, having an atomic origin, as discussed in the manuscript. The sample heating thus does not have an influence on our main conclusions and the background signal can be fully subtracted.

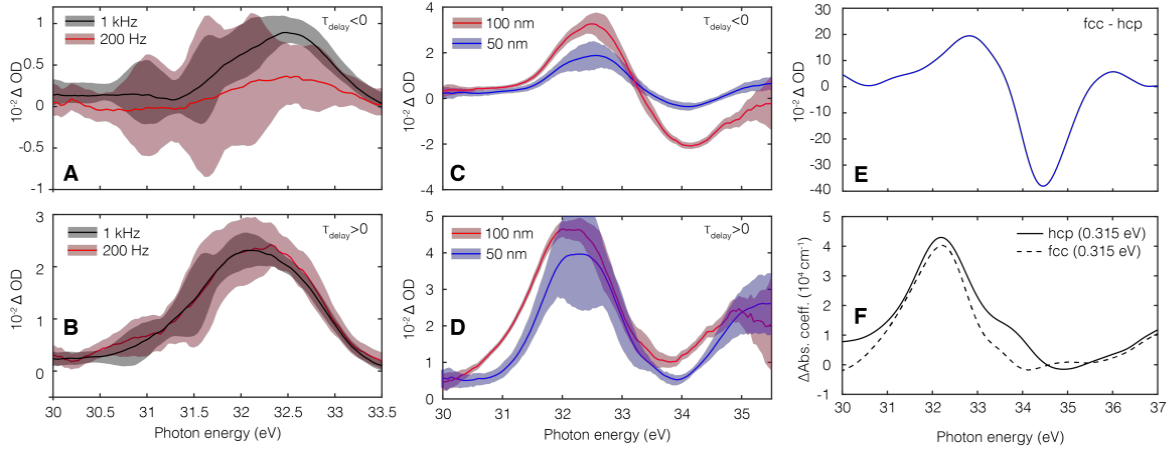

**Fig. S3**. **(A), (B)** Thermal background and transient signal, acquired in bare Ti films at different average powers (repetition rates). Shaded areas represent experimental uncertainties. Positive-delay profiles are already shown background-free. Only the thermal signal (A, negative delays) shows an average-power dependence. **(C), (D)** Sample thickness effect (bare 50-nm vs. 100-nm Ti foils) on thermal and transient signals. While the thermal signal is approximately proportional to sample thickness, the transient signal amplitude is not significantly affected. **(E)** TDDFT calculation of the optical density change due to hcp-to-fcc phase transition. Absorption of both phases calculated in the ground state. **(F)** Temperature-induced changes of hcp (solid line) and fcc (dashed) Ti phases, calculated with TDDFT. Both phases exhibit similar responses, confirming its universal and essentially atomic nature.

## Robustness of the induced spectral profile

We find that the observed pump-induced spectral profile of the optical density is robust with respect to film thickness (Fig. S3D) and oxidation condition (Fig. S4A). Figure S4A compares transient profiles of carbon-passivated and ambiently-oxidized titanium foils, at a pump-probe delay of ~ +10 fs (thermal background removed).

Moreover, an identical profile is detected if the sample is pumped with a long-wavelength source instead (commercial OPA source, ~30 fs, 1477 nm), as displayed in Fig. S4B. Figure S4C shows the long-wavelength pump spectrum, and Figs. S4D,E show corresponding transient absorption spectrogram and optical density dynamics (see Figs. 2A,C for comparison with 780-nm pump). This finding directly confirms that the transient absorption signal is of universal origin, and does not depend on the details of the excited electron distribution. Indeed, since the 3d orbitals span through the Fermi level, they can be populated either with 780-nm or 1477-nm pump photons. Therefore, ultrafast electron localization can be achieved in a wide range of photon energies, including the technologically important near- to mid-infrared range.

Finally, we observe the same structure in 100-nm thick Zirconium films (Fig. S4F), around the Zr $N_{2,3}$ edges of 27.1 and 28.5 eV, respectively[31]. The transition metal Zr belongs to the same group as Ti, therefore it is expected to show similar valence electron properties. Indeed, we find a similar structure above the $N_{2,3}$ absorption edge, which corresponds to 4p-4d transitions. This further confirms that ultrafast electron localization is a general phenomenon, enabled by the presence of available localized orbitals.

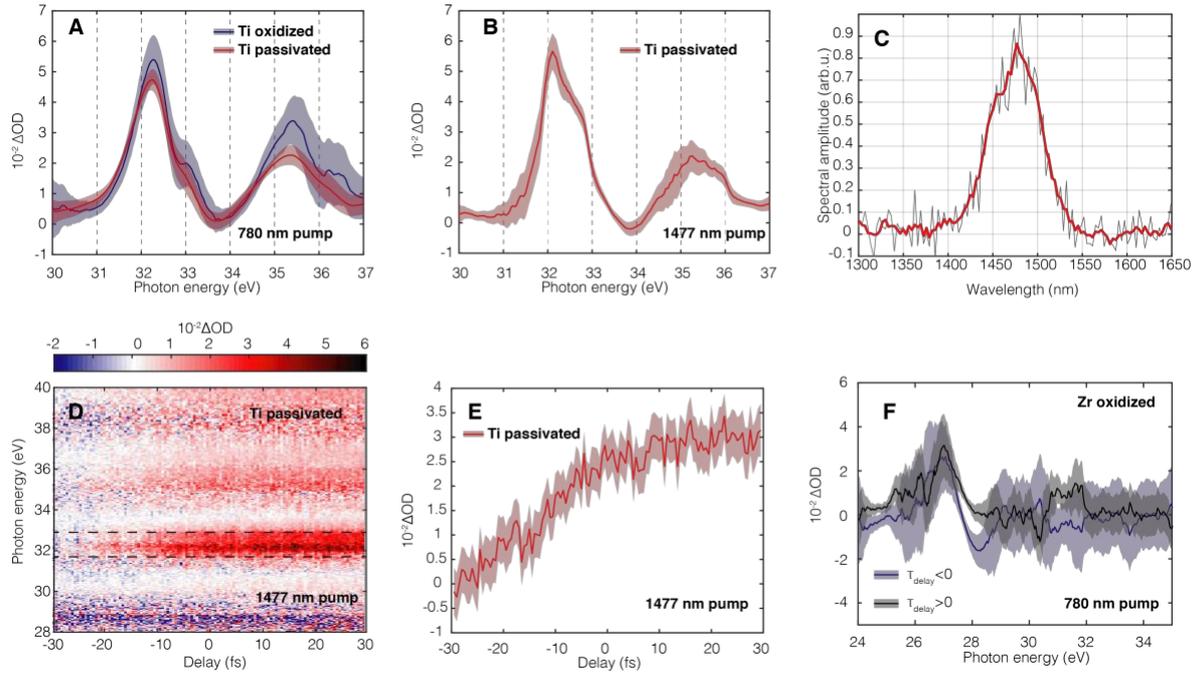

**Fig. S4**. **(A)** Transient absorption profiles of carbon-passivated Ti samples (red curve) compared with ambiently-oxidized Ti (blue curve). Shaded area represents the experimental standard deviation. **(B)** Transient profile of carbon-passivated Ti, pumped with a 1477-nm source, at a delay of ~+30 fs (thermal background removed). **(C)** Spectrum of the OPA source (black line), red line shows its smoothed average. **(D)** Transient absorption spectrogram, obtained with the 1477-pump, probed with APTs. (see Fig. 2A for comparison with 780-nm pump) **(E)** Corresponding optical density signal in the band of 31.7 eV to 32.9 eV (dashed lines in panel D). Shaded area represents the experimental standard error of the mean. **(F)** Transient and thermal profiles of optical density change in a 100-nm-thick Zr film. The Zr signal is probed with APTs, pumped with the regular 780-nm source. Shaded area shows the standard error of the mean.

## Response time extraction

To characterize the material response time with sub-cycle resolution, we reconstruct the temporal profile of the pump electric field via a photoelectron streaking experiment[49,50]. XUV-photoionized electrons are accelerated in the NIR pump field and collected in a time-of-flight spectrometer. Their kinetic energy as a function of XUV-NIR delay directly reflects the NIR field vector potential. In contrast to Refs. (*12, 14*), we perform the streaking experiment after the transient absorption measurement, with the metallic sample removed in order to avoid the electron background in the time-of-flight measurement. The actively stabilized pump-probe delay ensures the timing accuracy between the experiments down to a precision of 15 as rms.

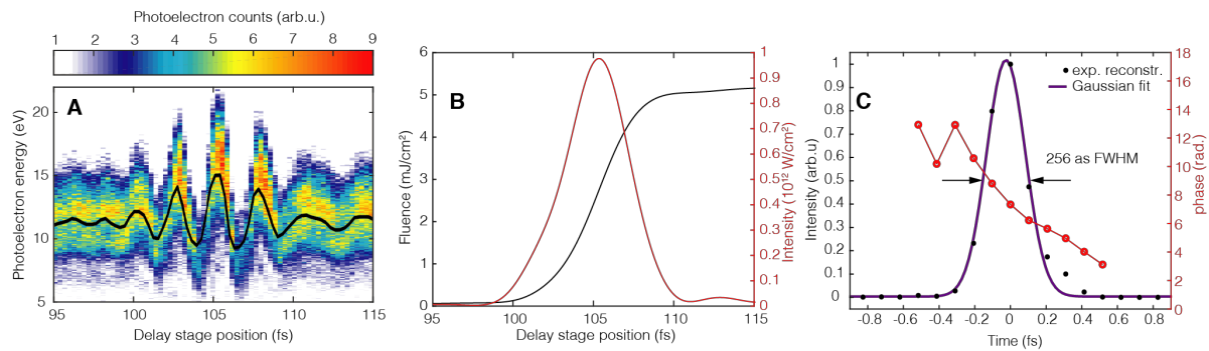

**Fig. S5.** **(A, B)** Photoelectron streaking (A) and NIR pulse intensity extracted from it (B, red curve). The black curve in B displays the associated IR fluence. **(C)** Probe pulse obtained with the ptychographic reconstruction

algorithm[51]. Violet line shows a Gaussian fit with 256 as FWHM (full width at half maximum). Red curve displays the reconstructed phase.

The laser intensity is defined as the absolute value of the inverse Fourier-transformed positive-frequency part of the vector potential, squared:

$$I_0(t) = \left(2|\widetilde{IF}(A_{\omega>0})|\right)^2, \qquad (S1.7)$$

where $A_{\omega>0}$ is the positive-frequency part of the Fourier-transformed vector potential, $\widetilde{IF}$ is the inverse Fourier-transform operator.

The vector potential convoluted with the XUV pulse duration is approximated by the center-of-mass of the experimental streaking trace:

$$A_{stk}(t) = A_{NIR}(t) \otimes I_{xuv}(t) \propto \frac{\int p N(p,t) dp}{\int N(p,t) dp}, \qquad (S1.8)$$

where $N(p,\tau)$ is the photoelectron count as a function of photoelectron momentum $p$ and pump-probe delay $t$. Since the transient absorption signal is also probed with the same pulse, we do not perform additional deconvolution of the vector potential. The proportionality coefficient is determined from independent measurements of the pulse energy and beam profile. The growth rate of the transient absorption signal, which we analyze in the band from 31.7 to 32.9 eV (Fig. 2A), is defined by a competition of non-equilibrated and thermalized contributions. We can however estimate the response time with a simple model, assuming a constant absorption coefficient for the whole pump spectral range.

We assume that the optical density response is linear and follows the absorbed laser intensity:

$$\Delta OD_{model}(t) = a \int_{-\infty}^{t} I_0(t' - t_{jitter}) h_\tau(t - t') dt' \qquad (S1.9)$$

with the response function having the following form:

$$h_\tau(t) = 1 - e^{-\frac{t}{\tau}}, \qquad (S1.10)$$

where $t_{jitter}$ accounts for a possible jitter between the streaking and transient absorption data due to acquisition noise, delay drift and spatial separation of the gas and solid targets.

We find the response time by simultaneously fitting 7 different pairs of streaking and transient absorption experiments, with parameters $a$ and $t_{jitter}$ being varied, and the response time $\tau$ being common for all the scans. This approach reflects the fact that $\tau$ is a constant of the system being studied, while $a$ and $t_{jitter}$ are attributable to experimental noise contributions. We minimize the sum of $\chi^2$ of 7 scans:

$$\{A\}, \{t_{jitter}\}, \tau : \min\left(\Sigma_{scan\,pairs}\,\chi^2\right),$$

$$\chi^2 = \Sigma_i \frac{(\Delta OD_{exp} - \Delta OD_{model})^2}{\sigma_i^2} \qquad (S1.11)$$

where $\Delta OD_{exp}$ is the experimental pump-induced optical density change in the energy band from 31.7 to 32.9 eV and $\sigma_\tau^2$ is the standard deviation of optical density (blue-shaded area in Fig.S6). We use the simplex-downhill algorithm to find the minimizing solution, starting with uniformly distributed initial guess parameters to check for local minima effects.

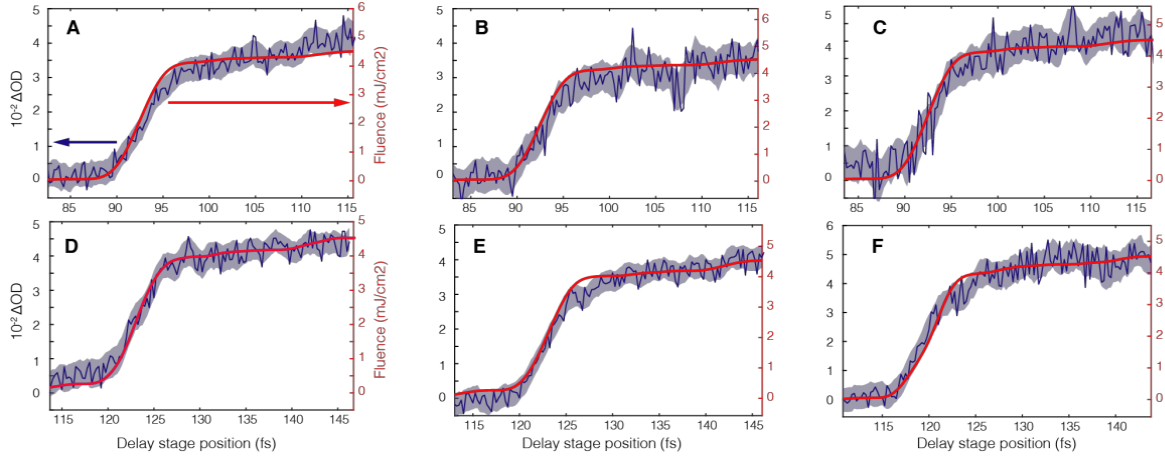

**Fig. S6.** Several representative transient absorption signals in the band from 31.7 to 32.9 eV, with corresponding laser fluence extracted from the subsequent streaking experiments, see also Fig. 2c.

To find the uncertainty of the response, we run a Monte-Carlo simulation. The experimental uncertainty on the optical density change is calculated as a standard error of the mean for 300 pairs of signal (pump on)/reference (pump off) spectra:

$$\sigma_{\Delta OD} = \sqrt{\frac{\sum_1^N (\Delta OD_i - \langle \Delta OD \rangle)^2}{N(N-1)}}, \Delta OD_i = \ln(I_i^{ref}/I_i^{sig}) \quad (S1.12)$$

where N is the number of signal/reference pairs per delay step. From this distribution, we generate 1000 transient absorption signals for each ATAS/STK pair, and run the fitting algorithm. The result of Monte Carlo simulations (Fig. S7A) gives a mean response time of 1.12 fs with a standard deviation of 290 as. However, we note that there is a correlation between the timing jitter and response time (Fig. S7C) that favors longer response times at the expense of negative jitter offsets (Fig. S7B). We have used a simple convolution model, which allows to estimate the linear microscopic response of the media, but neglects macroscopic effects of electron diffusion and probe pulse propagation through the sample. These processes may lead to a delayed macroscopic response, therefore the 1.12 fs value should be considered as an upper estimate of the microscopic response time.

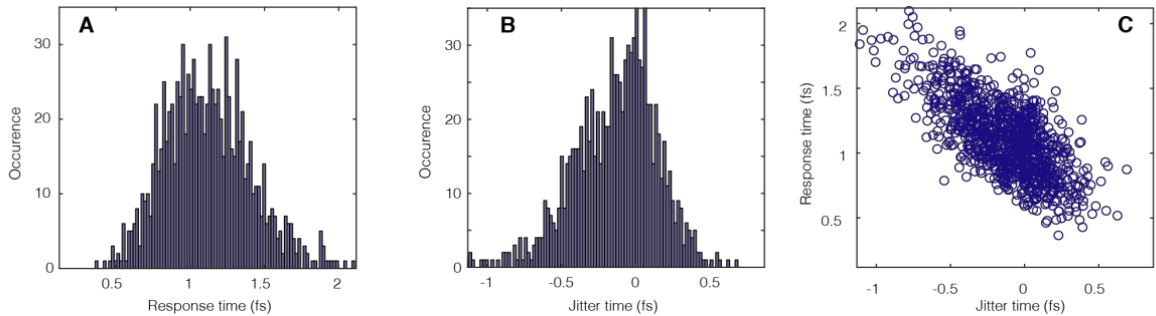

**Fig. S7**. **(A)** Histogram of response times, obtained with MC simulations. **(B)** An example (one of 7) histogram showing jitter time, as defined above. **(C)** Scatter plot showing anti-correlation of response and jitter times.

## S2. Theoretical modelling details

In order to understand the microscopic mechanism of the experimentally observed features in the attosecond transient absorption spectra, we employ *ab-initio* simulations based on the time-dependent density functional theory (TDDFT)[52]. In this section, we describe the TDDFT simulation and present further analysis based on it.

**Ab-initio electron dynamics simulation based on the TDDFT**

In the framework of the TDDFT, electron dynamics is described by a single-particle Schrödinger-like equation, which is the so-called time-dependent Kohn-Sham (TDKS) equation. For laser-induced electron dynamics in solids, the TDKS equation has the following form,

$$i\hbar \frac{\partial}{\partial t} u_{bk}(r,t) = h_{KS,k}(t) u_{bk}(r,t), \tag{S2.1}$$

where $u_{bk}(r,t)$ describe electron orbitals with band index $b$ and the Bloch wave vector $k$. The Hamiltonian $h_{KS,k}(t)$ is the so-called Kohn-Sham Hamiltonian, and is given by

$$h_{KS,k}(t) = \frac{1}{2m}\left(p + \hbar k + \frac{e}{c}A(t)\right)^2 + \hat{v}_{ion} + v_H(r,t) + v_{xc}(r,t), \tag{S2.2}$$

where $\hat{v}_{ion}$ is the ionic potential, $v_H(r,t)$ is the Hartree potential, and $v_{xc}(r,t)$ is the exchange-correlation potential. Note that the Hartee and exchange-correlation potentials are functionals of the electron density;

$$\rho(r,t) = \sum_{bk} n_{bk} |u_{bk}(r,t)|^2, \tag{S2.3}$$

where $n_{bk}$ is the occupation factor. In this work, we employ the adiabatic local density approximation (ALDA) for the exchange-correlation potential[53]. We describe the laser fields by a spatially uniform vector potential, $E(t) = -\frac{1}{c}\frac{dA(t)}{dt}$, assuming that the wavelength of the laser fields is much longer than the spatial scale of electron dynamics; this is nothing but the dipole approximation.

To describe the laser-induced electron dynamics in Ti metal, we employ the hexagonal close-packed (hcp) structure of Ti. Titanium atoms are described with a norm-conserving pseudopotential method, treating 3s, 3p, 3d, and 4s electrons as valence[54,55]. In the practical calculations, we solve the TDKS equation (S2.1) in real-space and real-time: the spatial coordinate **r** is discretized into uniform grid points with spacing h = 0.3 a.u., and the electron orbitals $u_{bk}(r,t)$ are propagated under the influence of the laser fields with a time step Δt = 0.03 a.u. The first Brillouin zone is also discretized into $20^3$ k-points. All the TDDFT simulations in this work are carried out with the *Octopus* code[56].

One of the most important outputs of the TDDFT calculations is the macroscopic current density, $J(t)$, which is directly related to the optical properties of solids. We compute the current by

$$J(t) = -\frac{e}{m\Omega} \int_\Omega dr \sum_{bk} n_{bk} [u_{bk}^*(r,t)\{p + \hbar k + eA(t)/c\} u_{bk}(r,t)] + J_{PS}, \tag{S2.4}$$

where $\Omega$ is the volume of the unit cell, and $J_{PS}$ is the contribution from the pseudopotential. If the electric current under an impulsive distortion, $E(t) = E_0 \delta(t)$, is computed, the optical conductivity of the system can be evaluated as

$$\sigma(\omega) = \frac{\int dt\, e^{i\omega t - \gamma t} J(t)}{E_0}, \tag{S2.5}$$

where $\gamma$ is a numerical damping factor. We set it to 0.5 eV/$\hbar$ in this work. From the optical conductivity, the absorption coefficient can be further computed.

The absorption coefficient in Fig. 1c is computed with TDDFT using this scheme. If the time-dependence of the Hartree and the exchange-correlation potentials is ignored in the time-propagation, this scheme provides the absorption coefficient with the independent particle (IP) approximation in Fig. 1c since the TDKS equation is simply reduced to the Schrödinger equation of an independent particle system.

**Optical response of pumped systems in non-equilibrium phase**
To investigate the modification of the absorbance immediately after the pump laser pulse, we perform a numerical pump-probe simulation based on the TDDFT (*34*). For the pump pulse, we employ the following form,

$$A_{pump}(t) = -\frac{cE_{pump}}{\omega_{pump}} \cos^2\left[\frac{\pi}{T_{pump}}t\right] \sin[\omega_{pump}t], \qquad (S2.6)$$

in the domain, $-T_{pump}/2 < t < T_{pump}/2$, and zero outside. Here $\omega_{pump}$ is the mean frequency, and $T_{pump}$ is the full duration of the pump pulse. We set $\omega_{pump}$ to 1.55 eV/$\hbar$, and $T_{pump}$ to 20 fs. The corresponding full-width half-maximum of the laser intensity is about 7 fs. We also set the maximum field strength $E_{pump}$ to $9.7 \times 10^8$ V/m, which corresponds to the maximum field strength at the front surface of the sample under an irradiation of $10^{12}$ W/cm$^2$, taking the surface reflection into account with the stationary solution of the Maxwell equation. For the probe pulse, we employ the impulsive distortion at the end of the pump pulse ($t = T_{pump}/2$).

To extract the optical property of the pumped system, we perform two TDDFT simulations. One employs both the pump and the probe pulses, while the other employs only the pump pulse. We then define two kinds of electric current. One is computed by the pump-probe calculation, while the other is computed by the pump-only calculation. We denote the first one *pump-probe current* $J_{pump-probe}(t)$ and the latter one *pump-only current* $J_{pump-only}(t)$. By subtracting $J_{pump-only}(t)$ from $J_{pump-probe}(t)$, one may extract the current induced by the probe pulse under the presence of the pump pulse. We shall call it *probe current,* $J_{probe}(t) = J_{pump-probe}(t) - J_{pump-only}(t)$. Applying the Fourier analysis of Eq. (S2.5) to the probe current, the optical property of the laser-excited system can be computed. As a result of this pump-probe scheme, the absorption coefficient in the non-equilibrium phase is provided (Fig. 3a).

**Optical response of pumped systems in electron-thermalized phase**
Immediately after the ultra-short laser pumping, the system is in a completely non-equilibrium phase. Then, the electronic system thermalizes and forms the hot-electron state within a 10-100 fs time scale (*10*). This phase can often be treated by a two-temperature model, where the electronic and the ionic systems are treated by thermal distributions with different temperatures; electron temperature and ion temperature. In order to theoretically investigate the effect of the electron thermalization, we compute the optical properties of hcp Ti with a finite electron temperature[57].
In practical calculations, we first compute the ground state, assuming the Fermi-Dirac distribution for the electron occupation $n_{bk}$ for the density in Eq. (S2.3). As a result, the initial electron orbitals $u_{bk}(r, t = 0)$ and the occupation $n_{bk}$ with finite electron temperature can be prepared consistently. Applying the impulsive distortion and computing the electron dynamics based on the above scheme, we then compute the optical properties of hcp Ti with finite electron temperature. In this work, we set the electron temperature $T_e$ to 0.315 eV so that the excess energy of the finite electron-temperature system becomes identical to that of the system pumped by the laser used in the above analysis. As a result of the TDDFT simulation with finite electron-temperature, the change in absorption coefficient in Fig. 3a is obtained.

**Decomposition of transient optical properties**
To clarify the microscopic mechanism of the absorption change induced by the pump pulse, we perform a theoretical decomposition of the computed transient absorption coefficient. Since the thermal and non-equilibrium electronic distributions provide qualitatively equivalent results, we focus on the thermal distribution for simplicity.

To construct the theoretical decomposition, we first revisit the linear response in the TDDFT with a weak electric field. In the weak perturbation limit, the KS Hamiltonian (S2.2) under the ALDA can be rewritten as

$$h_{KS,k}(t) = \frac{1}{2m}\left(\boldsymbol{p} + \hbar \boldsymbol{k} + \frac{e}{c}\boldsymbol{A}(t)\right)^2 + \hat{v}_{ion} + v_{Hxc}[\rho(\boldsymbol{r}, t=0)]$$
$$+ \int d\boldsymbol{r}' f_{Hxc}(\boldsymbol{r},\boldsymbol{r}')\delta\rho(\boldsymbol{r}',t), \qquad (S2.7)$$

where $v_{Hxc}[\rho(\boldsymbol{r}, t=0)]$ is the sum of the Hartree and the exchange-correlation potentials evaluated with the initial electron density $\rho(\boldsymbol{r}, t=0)$. The dynamical part of the Kohn-Sham potential can be evaluated with the Hartree-exchange-correlation kernel $f_{Hxc}(\boldsymbol{r},\boldsymbol{r}')$ and the induced density $\delta\rho(\boldsymbol{r}, t) = \rho(\boldsymbol{r}, t) - \rho(\boldsymbol{r}, t=0)$.

An essential difference between the optical response in the electron-thermalized phase and in the ground state is the choice of the occupation factor $n_{bk}$. In the present work, we employ the Fermi-Dirac distribution with finite temperature, $T_e = 0.315$ eV, for the electron-thermalized phase, while we employ almost zero temperature, $T_e = 0.01$ eV, for the ground state. Therefore, the change of the absorption due to the electron temperature can be analysed based on the occupation factor contribution. In the linear response TDDFT calculation, the occupation factor affects the following three parts: One is the electronic structure renormalization via the initial Hartree-exchange-correlation potential $v_{Hxc}[\rho(\boldsymbol{r}, t=0)]$ and the initial density $\rho(r, t=0)$. The second is the modification of the local screening effect via the modification of the induced potential and the induced density $\delta\rho(\boldsymbol{r}, t)$. The third is the state-filling effect via the current evaluation with Eq. (S2.4). To extract each contribution, we perform several simulations including each contribution alone. Figure S8 shows the change of the absorption coefficient of hcp Ti with the finite electron temperature, $T_e = 0.315$ eV. The violet line shows the full TDDFT result. Decomposed results are also shown: the band structure renormalization (red), state filing (blue), and the modification of the local screening effect (black). The reconstructed signal from the decomposed results is also shown as a black-dashed line. Since the reconstructed result shows nice agreement with the full TDDFT result, the validity of the decomposition can be confirmed. The tiny difference between the full TDDFT and the reconstructed results originates from the nonlinear coupling among the three contributions.

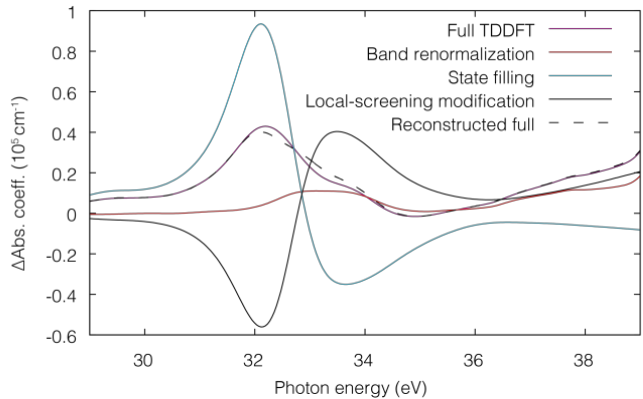

**Fig. S8.** Change of absorption coefficient computed with the finite electron-temperature TDDFT (violet). The result is decomposed into three contributions; band structure renormalization (red), state filling (blue), and modification of the local-screening effect (black). The reconstructed signal from the decomposed contributions is also shown as a black-dashed line.

## Quantification of electron-localization dynamics

To quantify the localization dynamics under a pump pulse, we compute the local induced density as

$$S(t) = \int dr\, [\rho(\mathbf{r}, t) - \rho(\mathbf{r}, t = 0)] e^{-(\mathbf{r}-\mathbf{R}_{Ti})^2/2\sigma^2}, \tag{S2.8}$$

where $\rho(\mathbf{r}, t)$ is the electron density under the pump pulse, and $\mathbf{R}_{Ti}$ is the position of a Ti atom. We set the width of the Gaussian window $\sigma$ to 1 a.u. The result is displayed in Fig. 4C.

## Nature of electronic localization

Since the 3d-shell of Ti is less than half-filled, the spatially confined 3d orbitals can be directly populated by optical excitation from 4s and 4p states, thus increasing the electronic localization. Occupation of 3d states is also influenced by subsequent electronic renormalization, since the Fermi level shifts to lower energies in Ti as a result of increased screening[35]. To pinpoint the origin of d-orbital population, we repeat the real-space simulations of pump-induced electronic density change with the independent-particle model, which does not allow for electronic renormalization (Fig. S9). The results again show an increase of electronic density in the shape of a d-orbital, thus showing that electronic localization is caused by their direct optical pumping.

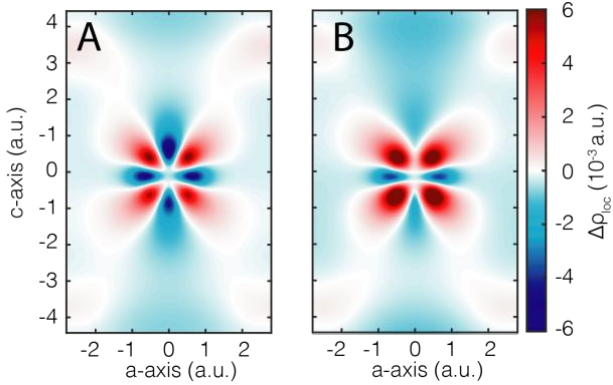

**Fig. S9**. Real-space, real-time simulation of the pump-induced electronic density change, right after the pump pulse. (A) shows the results from the full model, including the effects of electronic renormalization, while (B) shows the results in the IP (i.e. frozen potential) approximation. In both cases, the electron density localizes in the shape of d-orbitals, and thus mainly results from their direct optical pumping.

## On the Auger effect

The XUV-induced core-hole experiences an extremely fast relaxation via an intra-shell Auger process, known as super-Coster-Kronig decay. Due to strong interaction of localized and largely overlapping 3p and 3d states, inter-channel coupling leads to a strong modulation of the transition rate and may cause a failure of the single-particle core-hole decay picture[58,59]. Both the absorption and nonradiative decay are significantly enhanced at the Giant Resonance[8]. An excess population of d-bands therefore increases the decay probability, by means of increasing the number of participating Auger electrons and their interaction, thus shortening the excited state lifetime:

$$V_{eff} = \frac{V}{1 - \chi V} \tag{S.2.9}$$

$$\gamma \propto \sum_{i,f} w_i |\langle f | V_{eff} | i \rangle|^2 \delta(E_i - E_f) \tag{S.2.10}$$

where $V$ is inter-electron interaction, $\chi$ is the electron-vacancy propagator, $\gamma$ is the decay rate (see e.g. eq. (6.3) in reference[8]) and $w_i$ is the population of initial states. We can approximate the associated pump-induced effect on relaxation by line broadening. In Fig. S10, we have additionally broadened the absorption spectra of the ground state (Fig. S10A) and the finite electron temperature state (Fig. S10B) computed by the TDDFT. The additional broadening is set to 0.15 eV. The result reveals the second experimentally observed absorption peak at around 35 eV, though slightly shifted in energy. This discrepancy may be attributed to the many-body nature of the relaxation mechanism, which renders relaxation a dynamic process and causes a deviation from the simple line broadening.

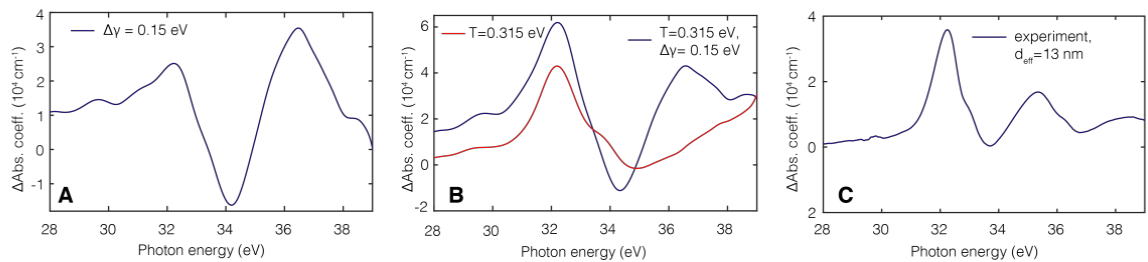

**Fig. S10.** Broadening effect on the XUV absorption spectrum. **(A)** The change of the absorption spectrum due to additional broadening to the static absorption from TDDFT. **(B)** Change of absorption spectra due to hot-electron (red line) and a combined hot-electron effect with empirical line broadening (blue line). **(C)** Absorption coefficient change, determined from the experiment assuming an effective 13 nm optical thickness due to the skin-effect.

**Band renormalization**

The band renormalization effect mainly originates from the 3p level shift, due to attenuation of the ionic core potential by excess screening of 3d-electrons. This effect is illustrated in Fig. S11, where we compare the TDDFT decomposition result with a 30 meV red-shift of the ground-state absorption spectrum.

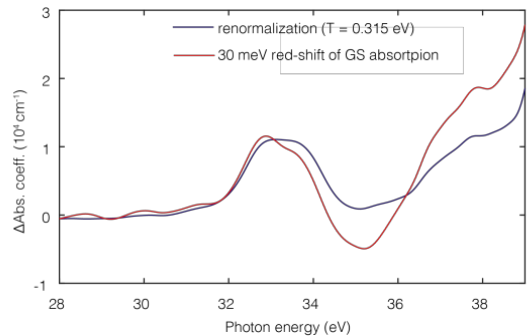

**Figure S11**. Band renormalization (blue) approximated as a red-shift of the ground-state (GS) absorption edge (red). The remaining differences between the two curves thus originate from renormalization of the valence electronic structure.

**References SI:**